\documentstyle[aps,twocolumn]{revtex}

\begin{document}

\twocolumn[\hsize\textwidth\columnwidth\hsize\csname
@twocolumnfalse\endcsname

\flushright{gr-qc/0001041 \\ OU-TAP 117 \\ YITP-00-16}

\title{ On cosmological perturbations in FRW model
with scalar field  \\
and false vacuum decay}

\author{Arsen Khvedelidze~$^{a,b}$, George Lavrelashvili~$^a$ and
  Takahiro Tanaka~$^{c,d}$\\$~$ }

\address{$^a$~Department of Theoretical Physics,
A.Razmadze Mathematical Institute, GE-380093 Tbilisi, Georgia}

\address{$^b$~Bogoliubov Laboratory of Theoretical Physics,
Joint Institute for Nuclear Research, Dubna, Russia}

\address{$^{c}$~Yukawa Institute for Theoretical Physics,
Kyoto University, Kyoto 606-8502, Japan}

\address{$^{d}$~Department of Earth and Space Science, Graduate School of
Science Osaka University, Toyonaka 560-0043, Japan}

\maketitle

\vspace{0.5cm}

\begin{abstract}
The unconstrained reduced action corresponding to the dynamics of
scalar fluctuations about the Friedmann-Robertson-Walker (FRW)
background is derived using Dirac's method of description of singular
Lagrangian systems.
The results are applied to so-called negative mode problem in description
of tunneling transitions with gravity.
With our special choice of physical variable, the kinetic term
of the reduced action has a conventional signature for a wide class of models.
In this representation, the existence of a negative mode justifying
the false vacuum decay picture turns out to be manifest.
We also explain how the present result becomes consistent with the
previously proved ``no negative mode (supercritical supercurvature mode) 
theorem''. 
\end{abstract}

\flushleft{PACS number(s): 98.80.Hw, 98.80.Cq}
 
\vspace{0.5cm}]

One of the interesting and widely exploiting cosmological models
is based on the theory of a scalar field coupled to gravity.
Such a remarkable phenomena as inflation and metastable (false) vacuum decay
are usually discussed in frame of this model.
For complete description of this processes, it is very important to know the
properties of perturbations to the background configurations describing an
inflation or vacuum decay. The cosmological perturbations in Lorentzian
regime are related to the cosmic microwave background radiation and
large scale structure formation \cite{MFB,GMST,GT}.
Furthermore the perturbations in the Euclidean version of the
theory define one loop corrections to the bubble nucleation rate and
determine quantum state of materialized bubble \cite{CC}.
Since the model contains gauge degrees of freedom,
the choice of physical variables to analyze the system
is not unique.

There are several known ways in deriving and expressing unconstrained
quadratic action solely written in terms of physical variables
in the theory of scalar field coupled to gravity
in non-spatially flat FRW Universe\cite{GMST,GT,LRT,TS,Lav-HT,TS2,Tanaka,GM}.
However, none of them is completely satisfactory for the purpose
to apply it to the issue of quntum tunneling, which we discuss mainly in
this short letter.
An extension of the convenient variable $v$ discussed in
the well-known review \cite{MFB} to the non-spatially flat
case is found in Ref.\cite{GM}.
However, the canonical transformation to arrive at this variable
turns out to be singular in the case of quantum tunneling.
In the reduction discussed in \cite{LRT}, the kinetic
term does not have definite signature.
In the case of the reduction given in \cite{TS,GMST}, the
kinetic term has a definite signature (as long as the
background scalar field is monotonic). However, the overall
signature is unconventional and hence some analytic continuation
similar to the conformal rotation becomes necessary.
The conformal rotation was introduced in the case of pure gravity
to cure the conformal factor problem\cite{GHP}.
It was also found to be possible to treat this problem
(at least in perturbative gravity)  via the careful
gauge fixing \cite{SH}.

In this short letter, we revisit the dynamics of small scalar-type
perturbations of a scalar field coupled to FRW type background in
the framework of conventional theory of degenerate Lagrangian
systems \cite{DiracL}. We obtain an expression for the
reduced action which is more appropriate for discussing the
quantum tunneling with gravity. For a wide class of models, the
signature of the obtained reduced action becomes the conventional
one. The potential for the eigen value problem is also regular for
the same class of models, and it is shown that there is a negative
mode in the spectrum of small fluctuations about the Coleman-De
Luccia bounce solution \cite{CD} in accordance with the consistent
interpretation of metastable vacuum decay \cite{CC}. As it will
become clear later, this result does not conflict with the
conclusions about the absence of negative mode arrived in the
other reduction scheme \cite{GMST,TS,Lav-HT,TS2,Tanaka}.

The  evolution  of system composed of scalar matter field minimally
coupled to gravity is determined by the conventional action
\begin{equation}
S =\int d^4x \sqrt{-g}\, \left[ \frac{1}{2\kappa} R - \frac{1}{2}\nabla_\mu
\phi\nabla^\mu\phi - V(\phi)\right]\; ,
\end{equation}
where $\kappa=8\pi G$ is the reduced Newton's constant and
$V(\phi)$ is the scalar field potential.

We expand the metric and the scalar field over an FRW type background
\begin{eqnarray} \label{eq:perturb}
 ds^2 & = & a(\eta)^2\Biggl[- (1 + 2A(\eta)Y)d\eta^2
   + 2{\cal B}(\eta) Y_{|i} d\eta dx^i \cr &&
 +\left\{\gamma_{ij}(1-2\Psi(\eta)Y)+2{\cal E}(\eta) Y_{|ij}\right\}
 dx^i dx^j\Biggr]          \nonumber,\\
 \phi & = & \varphi(\eta) + \Phi(\eta)Y,
\end{eqnarray}
where $\gamma_{ij}$ is the three-dimensional metric on the constant
curvature space sections, $a$ and $\varphi$ are the background field values
and $A, \Psi$, $\Phi$, ${\cal B}$ and ${\cal E}$ are small perturbations.
$Y$ is a normalized eigen function of 3-dimensional Laplacian,
$\Delta Y=-k^2 Y$,
and vertical line represents a covariant derivative with respect to
$\gamma_{ij}$.
To keep the simplicity, we set ${\cal B}(\eta)={\cal E}(\eta)=0$.
These terms are absent for homogeneous perturbations from the
beginning. Also for inhomogeneous perturbations, it is known
that this choice of gauge is consistent.

The background fields $a$ and $\varphi$ satisfy the equations
\begin{eqnarray}
{\cal H}^2 - {\cal H}' + {\cal K} &=& \displaystyle\frac{\kappa}{2}
  \,\varphi{}'{}^2 ,\\
2\, {\cal H}' + {\cal H}^2 + {\cal K} &=&
  \displaystyle \frac{\kappa}{2} \left(- \varphi'^2 +
                    2\, a^2 V(\varphi)\right), \\
\varphi{}'' + 2\, {\cal H} \varphi{}' + a^2
\displaystyle\frac{\delta V}{\delta \varphi} &=&0,
\end{eqnarray}
where a prime  denotes a derivative with respect to conformal time $\eta$,
${\cal H} := a'/a$, and ${\cal K}$ is the curvature parameter, which
has the values 1, 0, -1 for closed, flat and open universes, respectively.

Expanding the total action, keeping terms of second order in perturbations,
and using the background equations, we find
\begin{equation}
S = S^{(0)} + S^{(2)},
\end{equation}
where $S^{(0)}$ is the action for the background solution and
$S^{(2)}$ is quadratic in perturbations with the
Lagrangian for scalar perturbations
\begin{eqnarray}\label{eq:lgd}
^{(s)}{\cal L} = \frac{a^2\sqrt{\gamma}}{2 \kappa}&&\Biggl[
-6 \Psi'^2+2(k^2-3 {\cal K})\Psi^2 \cr &&
 + \kappa \left\{\Phi'^2 -
\left(a^2 \displaystyle\frac{\delta^2 V}{\delta\varphi\delta\varphi}
   +k^2\right) \Phi^2
+ 6 \varphi' \Psi' \Phi\right\}
         \nonumber \\
&&
 - \Bigl\{ 2\kappa\varphi' \Phi'
+2\kappa a^2 \displaystyle\frac{\delta V}{\delta \varphi} \Phi
\cr &&\quad\quad
+12 {\cal H} \Psi' +4(k^2-3{\cal K})\Psi \Bigr\} A \cr &&
 - 2 ({\cal H}' +
 2 {\cal H}^2- {\cal K} ) A^2 \Biggr].
\end{eqnarray}

To obtain the unconstrained system corresponding to the degenerate
Lagrangian (\ref{eq:lgd})
we will follow the Dirac's description of singular
Lagrangian  systems \cite{DiracL}.
Performing the Legendre transformation with canonical momenta
\begin{eqnarray}
\Pi_\Psi &:=&\displaystyle \frac{\delta ^{(s)}{\cal L}}{\delta \Psi'}
  = \frac{6\,a^2\,\sqrt{\gamma}}{\kappa}\left(-\Psi' +
   \frac{\kappa}{2}\varphi'\Phi - {\cal H}\, A \right) , \\
\Pi_{\Phi} &:=& \displaystyle \frac{\delta^{(s)}{\cal L}}{\delta \Phi'}
  = a^2\,\sqrt{\gamma}\left(\Phi' - \varphi'\, A\right) ,
\label{PiPhi}\\
\Pi_A &:=& \displaystyle \frac{\delta^{(s)}{\cal L}}{\delta A'} = 0 ,
\end{eqnarray}
we find the primary constraint \( C_1 := \Pi_A = 0 \).
Thus the total Hamiltonian $H_T$ is
\begin{equation}
H_T=H_C+ u_1 (\eta ) C_1 ,
\end{equation}
with arbitrary function \(u_1 (\eta )\) and canonical Hamiltonian
\begin{eqnarray}\label{eq:hamcan}
H_C &=& -\frac{\kappa}{12\,a^2\sqrt{\gamma}}\Pi_\Psi^2
+ \frac{1}{2\,a^2\sqrt{\gamma}}\Pi_\Phi^2 +
  \frac{\kappa}{2} \varphi' \,\Pi_\Psi\Phi
\cr&&
+ a^2\sqrt{\gamma} \Biggl[
 -\frac{k^2-3 {\cal K}}{\kappa} \Psi^2\cr &&
   + \frac{1}{2} \left(
   a^2 \displaystyle\frac{\delta^2 V}{\delta\varphi\delta\varphi}
    - \frac{3}{2}\kappa\varphi'^2
    +k^2 \right) \Phi^2 \Biggr]+ A \; C_2 \; ,
\end{eqnarray}
where
\begin{eqnarray}\label{eq:constraint}
C_2 &= & \varphi'\Pi_{\Phi} - {\cal H} \Pi_\Psi\cr &
+ &a^2 \sqrt{\gamma} \left[
\left(a^2\displaystyle\frac{\delta V}{\delta \varphi}
+ 3\varphi'{\cal H}\right) \Phi + \frac{2(k^2-3{\cal K})}
   {\kappa} \Psi \right] .
\end{eqnarray}
Conservation of primary constraint gives the secondary constraint
$C_2=0$. The primary and the secondary constraints are first class
and there are no ternary constraints. 

The existence of constraints in the system as usually
means the presence of gauge degrees of freedom.
To identify the physical degrees of freedom,
we fix the gauge and solve the constraints.
There are two simple strategies: either to eliminate
perturbations of scalar field ($\Pi_{\Phi}$ and $\Phi$) or
the gravitational degrees of freedom ($\Pi_{\Psi}$ and $\Psi$).
The approach developed in \cite{TS,Lav-HT,TS2,Tanaka,GMST} is
based on the first possibility.\footnote{Note that the gauge invariant
formulation in  \cite{GMST} corresponds to the gauge choice
$\chi_{\small GMST}:=\Phi=0$. }
Here we use the second possibility.

Thus we choose the following gauge fixing condition
\begin{equation}\label{eq:KLgauge}
\chi_1 := \Pi_\Psi = 0 \; . 
\end{equation}
{}From the consistency condition $\chi'_1=0$, we obtain 
\begin{equation} \label{Newton}
\chi_2=A-\Psi=0,
\end{equation}
which is the condition known as the Newton gauge. 
Then, the consistency condition $\Psi'- A'=0$ will 
determine $u_1$, and the set of constraints closes.
 

As a next step one can introduce the Dirac brackets,
or equivalently we can identify the Hamiltonian 
for the physical degrees of freedom 
$H^*$ by the relation
\begin{equation}
\Pi_{\Psi}\Psi'+\Pi_{\Phi}\Phi'+\Pi_A A'-
H_C \big\vert_{\chi_i=0, C_i=0}=
\Pi_{\Phi}\Phi'-H^*.
\end{equation}

After some algebra, assuming $k^2\neq 3{\cal K}$, we obtain
\begin{eqnarray} \label{eq:ham*}
 H^{*}&=&{{Q}\over 2a^2\sqrt{\gamma}} \Pi_{\Phi}^2 -
   {\kappa\varphi' \over 2(k^2-3{\cal K})}
   \left(3\varphi'{\cal H} +a^2 {\delta V\over \delta\varphi}\right)
   \Pi_{\Phi}\Phi\hspace{-5mm} \cr
  && +{1\over 2}a^2\sqrt{\gamma}\Biggl[
  -{\kappa\over 2(k^2-3{\cal K})}
   \left(3\varphi'{\cal H} +a^2 {\delta V\over \delta\varphi}\right)^2
 \cr && \quad\quad\quad\quad
+\left(a^2{\delta^2 V\over \delta\varphi\delta\varphi}
          -{3\over 2}\kappa\varphi'{}^2+k^2\right)
    \Biggr]\Phi^2,
\end{eqnarray}
with
\begin{equation}
  Q:= 1-{\kappa\varphi'^2 \over 2(k^2-3{\cal K})} \; .
\end{equation}

To proceed further it is useful to introduce new canonical coordinates
$f$ and $\pi_f$ defined by
\begin{eqnarray}\label{eq:cantrans1}
\pi_{f}
   & - &\left[Q{\cal H}+{\kappa\varphi'\over 6{\cal K} }{1+Q\over Q}
    \left(2\varphi'{\cal H} +a^2 {\delta V\over \delta\varphi}\right) \right]
     f\cr
 &= &\sqrt{Q\over a^2\sqrt{\gamma}}\Pi_{\Phi},  \\
f &= &\sqrt{a^2\sqrt{\gamma}\over Q}\Phi.     \label{eq:fdef}
\end{eqnarray}
One finds that the dynamics of physical variable $f $ is governed by
the simple harmonic oscillator Hamiltonian
\begin{equation}
{H}^\ast= \frac{1}{2}\pi^2_f  +
\frac{1}{2}w^2[a(\eta), \varphi(\eta)] f^2,
\end{equation}
with  frequency whose time dependence is determined by the background
solutions
\begin{equation}
w^2 [a(\eta), \varphi(\eta)] =
{\sqrt{Q}\over \varphi'}\left({\varphi'\over \sqrt{Q}}\right)''
      -(k^2 -3{\cal K}) Q.
\end{equation}
Thus, the unconstrained quadratic action becomes
\begin{equation}\label{eq:loract}
S^{(2)} =\int \left(
\frac{1}{2} f'^2 - \frac{1}{2} w^2[a(\eta), \varphi(\eta)] f^2
\right) d\eta \; .
\end{equation}

Now we apply the derived reduced action to investigation of the negative
mode problem in quantum tunneling with gravity.
The false vacuum decay is described by the bounce solution of Euclidean
equations\cite{Coleman,CD}. Value of the Euclidean action at the bounce
gives leading exponential factor in decay rate. Quadratic action defines one
loop corrections. It is remarkable that in the spectrum of small
perturbations about a bounce in absence of gravity there is exactly one
negative mode \cite{CC}.
This mode is responsible for making correction to the ground-state energy
imaginary, i.e., 
justifying decay interpretation.

Relevant object for tunneling transitions is the Euclidean
action which can be obtained from  the action Eq.~(\ref{eq:loract}) by the
analytic continuation  $\eta =- i \tau$. Defining the Euclidean action as
usual by $S^{(2)}= i S^{(2)}_E$ and specifying ${\cal K}=+1$,
we obtain
\begin{equation}\label{eq:eucact}
S^{(2)}_E =2 \pi^2 \int \left(
\frac{1}{2} \dot f^2 + \frac{1}{2} U [a(\tau),
     \varphi(\tau)] f^2 \right) d\tau  \; ,
\end{equation}
where $\dot{}=d/d\tau$ and
\begin{equation}
 U={\sqrt{Q_E}\over \dot\varphi}\left({d^2\over d\tau^2}
 {\dot\varphi \over \sqrt{Q_E}}\right)+(\ell(\ell+2)-3) Q_E,
\end{equation}
with
\begin{equation}
  Q_E= 1+{\kappa\dot\varphi^2 \over 2(\ell(\ell+2) -3)}.
\end{equation}
Here we used the fact that the eigen values of the Laplacian
$\Delta$ on a unit sphere take discrete values, i.e., $k^2=\ell(\ell+2)$ with
$\ell =0, 1, 2, ...$.

The equation for the mode functions, which diagonalize the action
Eq.(\ref{eq:eucact}), has form of the Schr\"odinger equation
\begin{equation}\label{eq:schrod}
-\frac{d^2}{d\tau^2} f + U [a(\tau), \varphi(\tau)]\, f
   = E\, f \; .
\end{equation}

Let's us first show that for $\ell=0$ case the
Eq.(\ref{eq:schrod}) has at least one negative mode for the
Coleman-De Luccia background bounce solution. 
We define a new potential
\begin{equation}
 \tilde U:={\sqrt{Q_E}\over \dot{\varphi}}
\left({d^2\over d\tau^2}{\dot{\varphi}\over \sqrt{Q_E}}\right) > U.
\end{equation}
The eigen value problem with this potential manifestly has
a zero eigen value state with $f\propto \dot{\varphi}/\sqrt{Q_E}$.
Since $\tilde U>U$, the eigen value problem with the potential $U$ must
have at least one negative eigen mode. $\Box$

A similar discussion leads to the conclusion
that there are no negative modes for $\ell>1$ (compare \cite{Lav-HT}).
The $\ell=1$ case needs separate consideration. As it was explained
in Ref.\cite{TS}, 
there are no physical degrees of freedom in this sector.

Note that the present result does not contradict with
the no negative mode theorem\cite{Tanaka}.
The argument of the no negative mode theorem
is that, when we consider the
bounce solution such that realizes the minimum value of action
among all the non-trivial $O(4)$-symmetric configurations,
there is no negative mode for the specific variable $q$ defined by
\begin{equation}\label{eq:qdef}
q:={8a\over \sqrt{\kappa}\dot\varphi}\Psi^N,
\end{equation}
where $\Psi^N$ represents $\Psi$ evaluated in the Newton gauge\cite{GMST}.
In terms of this variable $q$, the kinetic
term stays negative for $\ell=0$ modes.
Hence, as mentioned earlier,
we need to do some analytic continuation similar to the
conformal rotation to perform the path integral.
Although it is not fully justified,
this procedure of analytic continuation is thought to
produce the required imaginary factor. (See. \cite{TS,Tanaka}).
Now we are working with a different variable
$f$ for which the kinetic term takes the conventional
signature.
Therefore there should be one and only one 
negative mode.

Once we accept the original no negative mode theorem,
we can give an indirect proof of the uniqueness of
negative mode for $f$ under the same condition that
the absence of negative mode for $q$ was proven.
It is easy to find the relation between $f$ and $q$
because our gauge condition corresponds to the Newton gauge (\ref{Newton}).
Eliminating $\Pi_{\Phi}$ from the constraint equation $C_2=0$ and
the Hamiltonian equation of motion for $\Phi$ derived from
(\ref{eq:ham*}), we obtain 
\begin{equation}
{d\over d\tau}\left(
{\sqrt{Q_E}\over {}^4\sqrt{\gamma}\dot\varphi}f\right)
 ={ \ell(\ell+2) - {3{\cal K}}\over 4 \sqrt{\kappa}\dot\varphi} Q_E\, q,
\label{qtoPhi}
\end{equation}
where we used the definitions (\ref{eq:fdef}) and (\ref{eq:qdef}).
Suppose that $f$ has two negative modes for $\ell=0$,
and let us try to derive a contradiction by using the fact
that $q$ doesn't have any negative mode.
By assumption, the zero eigen value solution ($E=0$) of $f$ which
satisfies the boundary condition on one side must have two nodes.
One may think that this implies that there are at least two zeros of
$z:=({\sqrt{Q_E}/ {}^4\sqrt{\gamma}\dot\varphi})f$.
But it is not true because the regularity condition for $f$
just requires it to behave like $e^{-(\ell+1)|\tau|}$ on boundaries.
Since $\dot\varphi$ behaves like $\sim e^{-2|\tau|}$, even the regular
solution of $z$ does not go to zero on the boundary for $\ell=0$.
Hence, we arrive at the conclusion that there is at least one point
where the derivative of $z$ vanishes.
With the aid of Eq.(\ref{qtoPhi}), this implies
that the zero eigen value solution of $q$ has a node, which means
existence of a negative mode in $q$ and contradicts with the
no-negative mode theorem. $\Box$

We also note that with the present choice of variable there is no
manifest correspondence between no negative mode theorem
and the no supercritical supercurvature mode theorem \cite{TS2}
as was in terms of variable $q$ defined in \cite{GMST}.
The equation which determines the perturbation spectrum in the context
of one bubble open inflation in terms of the present variable $f$
is not standard Schr\"odinger-type equation and
the existence of a mode with a negative value of $E$ does not
imply the existence of a supercritical supercurvature mode.

So far we considered the background solutions with the
positive definite factor $Q_E$.
If this factor becomes zero or negative for some region(s) of  $\tau$,
the Euclidean analog of the canonical transformation
Eqs.(\ref{eq:cantrans1}), (\ref{eq:fdef}) becomes singular.
This will not immediately indicate a certain physical meaning
because there happens nothing special as long as
we discuss in terms of $q$ in Ref.\cite{GMST}.
However, if there exist some class of bounce solutions for which
the signature of kinetic term cannot be set to be
positive definite without passing through a singular
canonical transformation, then it might suggest
some physical meaning.
In this case, we might have to reconsider the possibility
of catastrophic particle creation discussed in Ref.\cite{LRT}.

To conclude we have investigated the dynamics of small
perturbations in non-flat FRW model coupled to a 
scalar field.
Using the gauge conditions Eq.(\ref{eq:KLgauge})
we reduced the system of coupled perturbations
Eq.(\ref{eq:lgd}) to the dynamical system Eq.(\ref{eq:loract})
with one physical degree of freedom.
The reduced quadratic action Eq.(\ref{eq:loract}) has the
conventional overall signature.
Investigating Euclidean quadratic action Eq.(\ref{eq:eucact}),
we proved that there is exactly one negative eigen value mode about the
Coleman-De Luccia bounce solution.
The result is consistent with the so-called  ``no-negative mode"
theorem \cite{Tanaka} in false vacuum decay with gravity
which was proven in the other reduction scheme.
The treatment discussed here is restricted to the background solutions
which satisfy the condition that the quantity $Q_E$ is positive.
Hence, another question arises 
whether we can always find a variable for
which the kinetic term of a perturbation in the reduced action
becomes positive definite.
This issue needs further investigation.

\acknowledgments
G.L. is grateful to D. Maison, V.A. Rubakov and G. Tsitsishvili
for many interesting discussions and comments at different stages of this
work.
T.T. thanks J. Garriga and X. Montes for valuable comments.
Work of G.L. was partly supported by the Georgian Academy of Sciences
under Grant No 1.4 and by the Royal Society Grant.
Work of T.T. was partly supported by
Monbusho System to Send Japanese Researchers Overseas.

\end{document}